\documentclass[prb,twocolumn]{revtex4}
\usepackage{graphicx}
\usepackage{amsmath}
\usepackage{amssymb}
\usepackage{latexsym}

\begin{document}

\title{Effects of boundary roughness on a $Q$-factor of whispering-gallery-mode lasing microdisk cavities}

\author{A. I. Rahachou and I. V. Zozoulenko}
\affiliation{Department of Science and Technology (ITN), Link\"oping University, 601 74
Norrk\"oping, Sweden}
\date{\today}

\begin{abstract}
We perform numerical studies of the effect of sidewall
imperfections on the resonant state broadening of the optical
microdisk cavities for lasing applications.  We demonstrate that
even small edge roughness ($\lesssim \lambda/30$) causes a drastic
degradation of high-$Q$ whispering gallery (WG) mode resonances
reducing their $Q$-values by many orders of magnitude. At the same
time, low-$Q$ WG resonances are rather insensitive to the surface
roughness. The results of numerical simulation obtained using the
scattering matrix technique, are analyzed and explained in terms
of wave reflection at a curved dielectric interface combined with
the examination of  Poincar\'{e} surface of sections in the
classical ray picture.
\end{abstract}
\maketitle

During recent years significant experimental efforts were put
forward towards investigation  of laser emission of dielectric and
polymeric low-threshold microdisk cavities
\cite{Introduction,Nockel,Slusher_1992,Berggren,Fujita,Gayral,Seassal,Inganas,Polson2002}.
The high efficiency of lasing operation in such devices is related
to the existence of natural cavity resonances known as whispering
gallery (WG) modes. The origin of these resonances can be
envisioned in a ray optic picture, wherein light is trapped inside
the cavity through total internal reflections on the cavity-air
boundary.

One of the most important characteristics of cavity resonances is
their quality factor ($Q$-factor) defined as $Q=2\pi*$(Stored
energy)/(Energy lost per cycle). The $Q$-factor of a microdisk
cavity is mostly governed by a radiative leakage through the
curved interface due to diffraction. An estimation of the
$Q$-factor in an ideal disk cavity of a typical diameter $d\sim
10\mu$m for a typical WG resonance gives $Q\sim 10^{13}$ (see
below, Eq. (\ref{Eq_T})).
At the same time, experimental measured values reported so far are
typically in the range of $10^3\sim 10^4$ or lower. A reduction of
a $Q$-factor may be attributed to a variety of reasons including
side wall geometrical imperfection, inhomogeneity of the height
and diffraction index of the disk, effects of coupling to the
substrate or pedestal and others. Several experimental studies
point out side wall imperfections as the main factor affecting the
$Q$-factor of the cavity\cite{Fujita,Gayral,Seassal}. An indirect
indication of the importance of this factor in disc microcavities
is provided by the observation that typical $Q$-factors of
\textit{spheroidal} microcavities are several orders of magnitude
higher than those of microdisk of comparable
dimensions\cite{Introduction,Ilchenko}. This is believed to be due
to superior quality of the microsphere surfaces where boundary
scattering may be limited by thermal fluctuations of the surface
only. Therefore, the effect of surface roughness appears to be of
crucial importance for the design, tailoring and optimization of
$Q$-values of lasing mikrodisk cavities. To the best of our
knowledge, this effect has not been considered to date and
warrants an investigation.


In order to compute the resonant states of a cavity of an
arbitrary shape we develop a new approach based on the scattering
matrix technique. The scattering matrix technique is widely used
in analyse of waveguides\cite{Russian} as well as in quantum
mechanical simulations\cite{Datta}.  This technique was also used
for an analysis of resonant cavities for geometries when the
analytical solution was available\cite{Hentschel}. Note that
because the problem at hand requires a fine discretization of the
geometry, commonly used time-domain finite difference
methods\cite{Li} would be prohibitively expensive in terms of both
computer power and memory. While a detailed description of the
calculations will be given elsewhere, we present  here the essence
of the method.

We consider a two-dimensional cavity with the refraction index $n$
surrounded by air. Because the majority of experiments are
performed only with the lowest transverse mode occupied, we
neglect the transverse ($z$-) dependence of the field and thus
limit ourself to the two-dimensional Helmholtz equation. We divide
our system in an outer and an inner regions. In the outer region
the refraction index $n$ is independent of the coordinate and the
solution to the Helmholtz equation can be written in polar
coordinates in the form
 \begin{equation}\label{psi_out}
 \Psi=\sum_{q=-\infty}^{+\infty}\left(A_q H_q^{(2)}(kr)+B_q H_q^{(1)}(kr)
 \right)e^{iq\varphi},
 \end{equation}
$\Psi=E_z\ (H_z)$ for TM (TE)-modes, $H_q^{(1)},H_q^{(2)}$ are the
Hankel functions of the first and second kind of order $q$
describing respectively incoming and outgoing waves,
$k=\omega/c=2\pi/\lambda$.

We define the scattering matrix $\mathbf{S}$ in a standard
fashion\cite{Russian,Datta,Hentschel}, $B=\mathbf{S}A$,
where $A, B$ are column vectors composed of the expansion
coefficients $A_q,B_q$ for incoming and outgoing states in Eq.
(\ref{psi_out}). The matrix element $\mathbf{S}_{q'q}$ gives the
probability amplitude of scattering from an incoming state $q$
into an outgoing state $q'$. In order to apply the scattering
matrix technique we divide the inner region into $N$ narrow
concentric rings. At each $i$-th boundary between the rings we
introduce the scattering matrix $\mathbf{S^i}$ that relates the
states propagating (or decaying) towards the boundary, with those
propagating (or decaying) away of the boundary. The matrices
$\mathbf{S^i}$ are derived using the requirement of continuity of
the tangential components for the $E$- and $H$-field at the
boundary between the two dielectric media. Successively combining
the scattering matrixes for all the
boundaries\cite{Russian,Datta},
$\mathbf{S^1}\otimes\ldots\otimes\mathbf{S^{N}}$, we can relate
the combined matrix to the scattering matrix $\mathbf{S}$.

To identify the resonant states of a resonant cavity we introduce
the Wigner-Smith time-delay matrix \mbox{$\mathbf{Q}=\frac{i}{c}\,
(d\mathbf{S^\dag}/dk)\,\mathbf{S}$}, \cite{Smith,Nockel,Hentschel}
where  the diagonal elements $\mathbf{Q}_{qq}$ give a time delay
experienced by the wave incident in $q$-th channel and scattered
into all other channels. The $Q$-value of the cavity is $Q=\omega
\tau_D(k)$, where $\tau_D(k)$ is the total time delay averaged
over all $M$ incoming channels \cite{Mello,Nockel,Hentschel},
\begin{eqnarray}
\tau_D(k)=\frac{1}{M}\sum_{q=1}^M\mathbf{Q}_{qq}=
\frac{1}{cM}\sum_{\mu=1}^{M}\frac{d \theta_\mu
}{dk}=\frac{1}{cM}\frac{d \theta}{dk}\, ,
\end{eqnarray}
$\exp(i\theta_\mu)=\lambda_\mu$ are the eigenvalues of the
scattering matrix $\mathbf{S}$, $\theta=\sum_{\mu=1}^N\theta_\mu$
is the total phase of the determinant of the matrix $\mathbf{S}$,
$\det \mathbf{S}=\prod_{\mu=1}^{M}\lambda_\mu=\exp(i\theta)$.

\begin{figure}[!htp]
\includegraphics[scale=0.3]{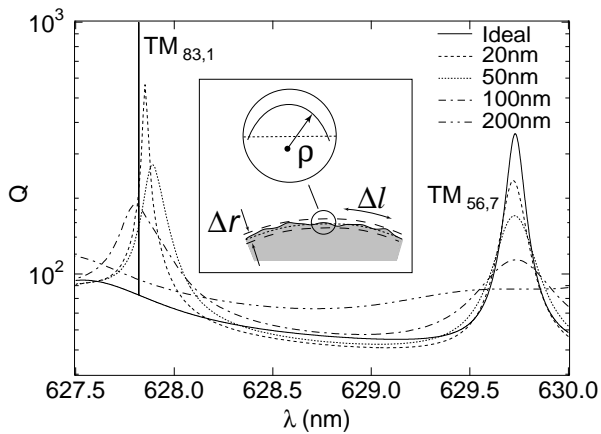}
\caption{(a) Dependence of the quality factor $Q=Q(\lambda)$ of
the circular disk for different surface roughness $\Delta r$
indicated in the figure; the disk radius is $R=5 \mu$m, the
refraction index $n=1.8$. The inset illustrates a cavity where the
surface roughness $\Delta r = 200$nm and $\Delta l=2\pi R/50$ (the
dotted line represents an ideal circular boundary, the shaded
region corresponds to the cavity). $\rho$ characterizes the
average radius of local curvature due to boundary imperfections.
TE modes of the cavity exhibit similar features and are not shown
here.} \label{spectra}
\end{figure}

Figure \ref{spectra} shows calculated $Q$-values of the disk
resonant cavity for  different surface roughnesses for TM modes in
some representative wavelength interval. Note that an exact
experimental shape of the cavity-surface interface is not
available. We thus model the interface shape as a superposition of
random Gaussian deviations from an ideal circle of radius $R$ with
a maximal amplitude $\Delta r/2$ and a characteristic distance
between the deviation maxima $\Delta l\sim 2\pi R/50 $ (see
illustration in inset to Fig. \ref{spectra}).

The solid curve in Fig. \ref{spectra} corresponds to an ideal disk
cavity without imperfections. Resonant states of an ideal disk (as
well as the bound states of the corresponding closed resonator)
are characterized by two numbers, $q$ (see Eq. (\ref{psi_out}) and
$m$. The index $m$ is a radial wave number that is related to the
number of nodes of the field components in the radial direction
$r$. The angular wave number $q$ can be related to the angle of
incidence $\chi$ in a classical ray picture \cite{Nockel}
\begin{equation}
q=nkR\, \sin\chi .\label{semiclassics}
\end{equation}

\begin{figure}[!htp]
\includegraphics[scale=0.3]{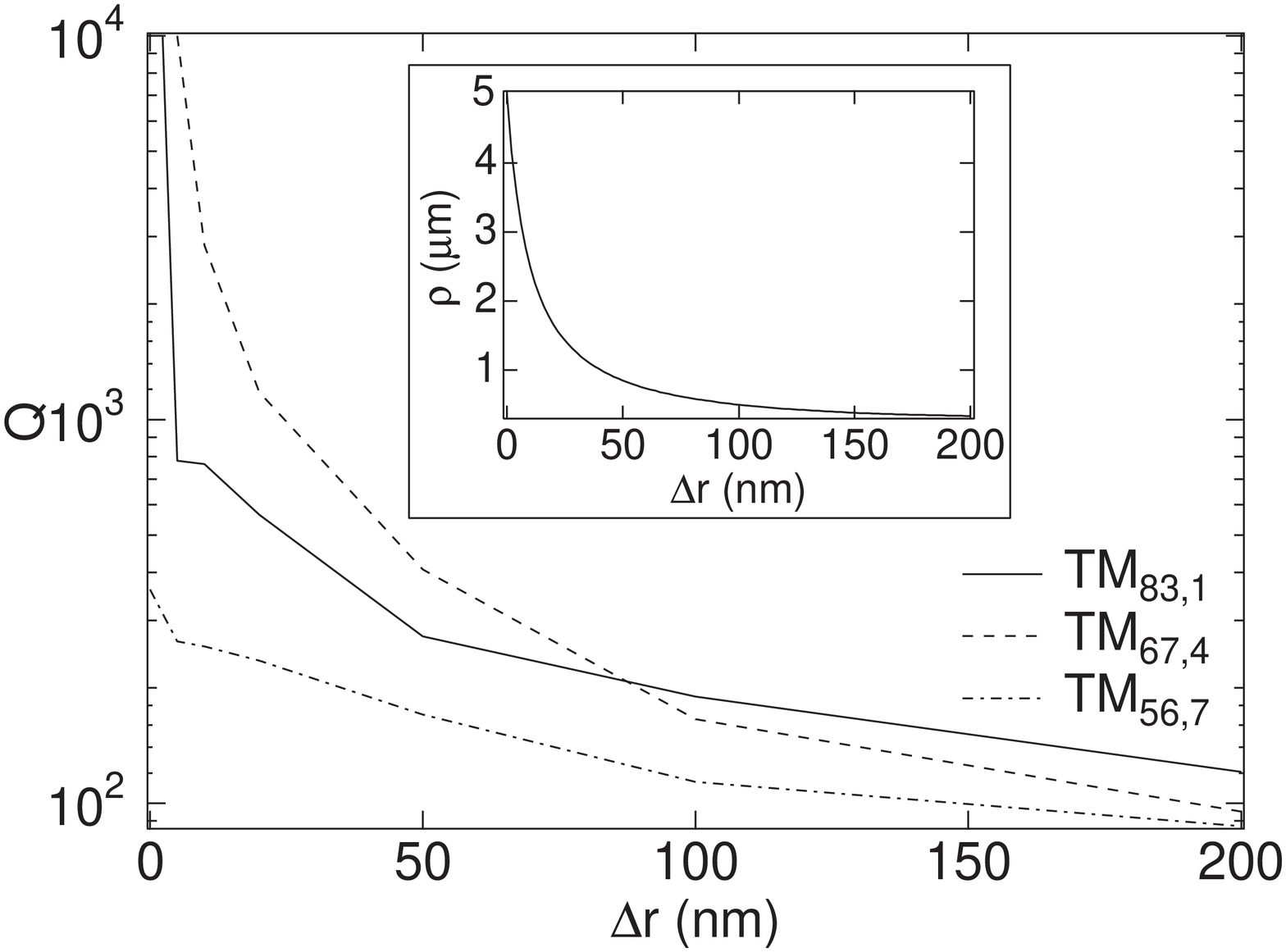}
\caption{Dependence $Q$ on the surface roughness $\Delta r$ for
several representative resonances. (Each curve remains practically
unchanged for different realizations of surface roughness). The
inset shows the dependence of local radius of roughness curvature
$\rho$ subject to $\Delta r$. Parameters of the cavity are the
same as in Fig. \protect\ref{spectra}. } \label{Q}
\end{figure}

The dependence of the averaged $Q$-values on the surface roughness
$\Delta r$ is summarized  in Fig. \ref{Q} for several
representative resonances. A common feature of all high-$Q$
resonances is a dramatic decrease of their maximal $Q$-value that
takes place for very small values of $\Delta r\lesssim\lambda/20$.
For example, a $Q$-value of at he resonant state TM$_{83,1}$ drops
from $Q\approx 10^{13}$ for an ideal disk  to $Q\approx 10^3$ for
surface roughness of only $\Delta r=20$nm.
In contrast, low-$Q$ resonances show a rather slow decrease in
their $Q$-values over the range of variation of $\Delta r$. For
example, for the same surface roughness $\Delta r=20$ the
$Q$-value of the resonant state TM$_{56,7}$ decreases only by a
factor of 1.5, dropping to $Q\approx 200$.

\begin{figure}[!htp]
\includegraphics[scale=0.3]{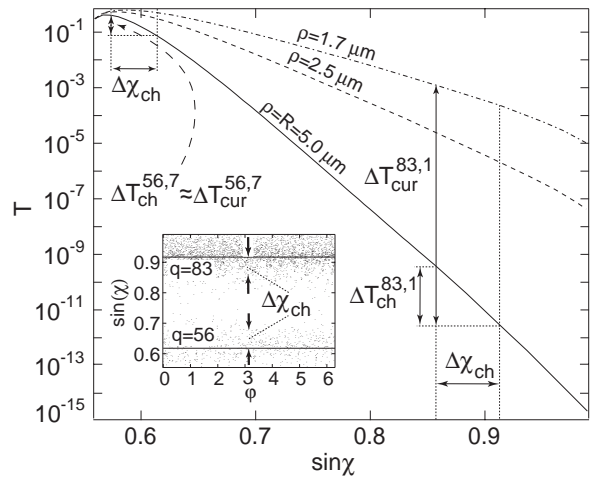}
\caption{Dependence $T=T(\chi)$  for several radii of curvature
$\rho$ according to Eq. \ref{Eq_T}. Inset shows a Poincar\'{e} SoS
for the states $q=83$ and $q=56$ for the cavity with $\Delta r =
0$ (straight lines of $\chi=\textrm{const}$) and $\Delta r =
20$nm. The number of bounces for a given angle of incidence
$\chi_0$ is chosen in such a way that the total path of the ray
does not exceed the one extracted from the numerically calculated
$Q$-value for the corresponding resonance, $L=c\tau_D=Q/k$.}
\label{T}
\end{figure}

In order to understand these features we combine
 a Poincar\'{e}  surface of section (SoS) method with an analysis of ray
reflection at a curved dielectric interface \cite{Snyder}. The
$Q$-value of the cavity can be related to the transmission
probability $T$ of an electromagnetic wave incident  on a curved
interface of radius $\rho$ by $Q=2nk\rho \cos\chi /T$
\cite{Hentschel_rapid} (this expression is valid for large angles
of incidence $\chi$ when $T\ll 1$). In turn, for $kn\rho \gg 1$,
the transmission probability reads \cite{Snyder}
\begin{eqnarray}\label{Eq_T}
T = |T_{F}| \exp{\left[-\frac{2}{3}\frac{nk\rho}{\sin^2(\chi)}
\left(\cos^2\chi_{c}-\cos^2\chi\right)^{3/2}\right]},
\end{eqnarray}
where $T_{F}$ is the classical Fresnel transmission coefficient
for an electromagnetic wave incident on a flat surface,
$\chi_{c}=\arcsin(1/n)$ is an angle of total internal reflection.
Figure \ref{T} illustrates that $T$ decreases exponentially as the
difference $\chi-\chi_c$ grows.

The inset to Fig.  \ref{T} depicts the Poincar\'{e} SoS for two
states with $q=56$ and $83$ shown in Fig. \ref{spectra}, where the
initial angle of incidence $\chi_0$ of launched rays is related to
the angular number $q$ by Eq. (\ref{semiclassics}). The SoS
demonstrates that initially regular dynamics of an ideal cavity
transforms into a chaotic one even for a cavity with maximum
roughness $\Delta r \lesssim 20$nm.
$\Delta T_{\textrm{ch}}^{83,1}$ in Fig. \ref{T} indicates the
estimated increase in the transmission coefficient due to the
broadening of the phase space, $\Delta \chi_{\textrm{ch}}$, as
extracted from the Poincar\'{e} SoS  for the state with $q=83$.
This corresponds to the decrease of $\Delta Q\sim\Delta
T^{-1}\approx 10^{-2}$. This value is much smaller that the actual
calculated decrease of the $Q$-factor for the high-$Q$ resonance
TM$_{83,1}$.

To explain the rapid degradation of high-$Q$ resonances, we
concentrate on another aspect of the wave dynamics. Namely, the
imperfections at the surface boundary effectively introduce a
local radius of surface curvature $\rho$  that is distinct from
the disk radius $R$ (see illustration in Fig. \ref{spectra}). One
may thus expect that with the presence of a local surface
curvature, the total transmission coefficient will be determined
by the averaged value of $\rho$ rather than by the disk radius
$R$. The dependence of $\rho$ on surface roughness $\Delta r$ for
the present model of surface imperfections is shown in the inset
to Fig. \ref{Q}. Figure \ref{T} demonstrates that  the reduction
of the local radius of curvature from $5\mu$m (ideal disk) to
$1.7\mu$m ($\Delta r = 20$nm) causes an increase of the
transmission coefficient by $\Delta T_{\textrm{cur}}\approx 10^8$.
This estimate, combined with the estimate based on the change of
$\Delta T_{\textrm{ch}}$ is fully consistent with the $Q$-factor
decrease shown in Figs. \ref{spectra},\ref{Q}. We thus conclude
that the main mechanism responsible for the rapid degradation of
high-$Q$ resonances in non-ideal cavities is the enhanced
radiative decay through the curved surface because the effective
local radius (given by the surface roughness) is smaller that the
disk radius $R$.

For the case of low-$Q$ resonances the change in the transmission
coefficient due to enhanced radiative decay $\Delta
T_{\textrm{cur}}$ is of the same magnitude as the change $\Delta
T_{\textrm{ch}}$ due to the broadening of the phase space caused
by the transition to chaotic dynamics  (as illustrated in Fig.
\ref{T} for the resonance TM$_{56,7}$). Therefore, both these
factor play comparable roles in degradation of the low-$Q$ WG
resonances.

It is worth mentioning that one often assumes that long-lived
high-$Q$ resonances in idealized cavities (e.g. in ideal disks,
hexagons, etc.) are not important for potential application in
optical communication or laser devices\cite{Hentschel,Wiersig}
because of their extremely narrow width. Our simulations
demonstrate that it is not the case, because in real structures
 the $Q$-values of these
resonances becomes comparable to those of intermediate-$Q$
resonances already for small or moderate surface roughness of
$\Delta r\sim 10-50$ nm.

To conclude, our results highlight the importance of surface
roughness for the performance of microcavities for laser
applications, and provide estimations on surface roughness that
might be instrumental for device design and fabrication.

We thank Olle Ingan\"as for stimulating discussions that initiated
this work and we thankful to Stanley Miklavcic and Sayan Mukherjee
for many useful discussions and conversations. A.I.R. acknowledges
financial support from SI and KVA.

\end{document}